\documentclass[11pt]{article}

\usepackage[preprint]{acl}

\usepackage{times}
\usepackage{latexsym}

\usepackage[T1]{fontenc}

\usepackage[utf8]{inputenc}
\usepackage{booktabs}

\usepackage{microtype}

\usepackage{inconsolata}

\usepackage{graphicx}

\usepackage{amsmath}

\usepackage{stfloats}
\usepackage{multirow} 

\usepackage{pgfplots}
\pgfplotsset{compat=1.18}
\usepackage{tikz}
\usetikzlibrary{matrix,arrows.meta,positioning,fit,calc,backgrounds}
\usepackage{xcolor}
\usepackage{amssymb}
\usepackage[most]{tcolorbox}
\usepackage{fvextra}

\usepackage{enumitem}

\usepackage{makecell}

\usepackage{seqsplit}

\usepackage{pgfplots}
\pgfplotsset{compat=1.18} 

\usepackage{xcolor}
\usepackage{booktabs}
\usepackage[edges]{forest}
\usepackage{adjustbox}
\usepackage{threeparttable}

\usepackage[flushmargin]{footmisc}

\newcommand{\code}[1]{\nolinkurl{#1}}
\makeatletter
\newcommand{\tabrowmark}[2]{%
  \expandafter\gdef\csname tabrow@#1\endcsname{#2}%
  \hypertarget{tabrow:#1}{\textup{(#2)}}%
}
\newcommand{\tabrow}[1]{%
  \hyperlink{tabrow:#1}{\textup{(\csname tabrow@#1\endcsname)}}%
}
\makeatother

\newtcolorbox{appendixsetupbox}[1]{
  enhanced,
  breakable,
  colback=blue!3,
  colframe=blue!55!black,
  colbacktitle=blue!55!black,
  coltitle=white,
  title=\textbf{#1},
  fonttitle=\bfseries,
  boxrule=0.5pt,
  arc=2pt,
  left=3pt,
  right=3pt,
  top=3pt,
  bottom=3pt
}

\title{Rethinking Agentic Search with \textsc{Pi-Serini}:\\Is Lexical Retrieval Sufficient?}

\author{ \textbf{Tz-Huan Hsu} \\ University of Waterloo \And \textbf{Jheng-Hong Yang} \\ Stencilzeit \And \textbf{Jimmy Lin} \\ University of Waterloo}

\begin{document}
\maketitle

\begin{abstract}
Does a lexical retriever suffice as large language models (LLMs) become more capable in an agentic loop? 
This question naturally arises when building deep research systems.
We revisit it by pairing BM25 with frontier LLMs that have better reasoning and tool-use abilities.
To support researchers asking the same question, we introduce \textsc{Pi-Serini}, a search agent equipped with three tools for retrieving, browsing, and reading documents.
Our results show that, on BrowseComp-Plus, a well-configured lexical retriever with sufficient retrieval depth can support effective deep research when paired with more capable LLMs.
Specifically, \textsc{Pi-Serini} with \code{gpt-5.5} achieves 83.1\% answer accuracy and 94.7\% surfaced evidence recall, outperforming released search agents that use dense retrievers.
Controlled ablations further show that BM25 tuning improves answer accuracy by 18.0\% and surfaced evidence recall by 11.1\% over the default BM25 setting, while increasing retrieval depth further improves surfaced evidence recall by 25.3\% over the shallow-retrieval setting.
Source code is available at~\url{https://github.com/justram/pi-serini}.
\end{abstract}

\section{Introduction}

Retriever effectiveness is widely treated as the hard ceiling of information-seeking systems such as Retrieval-Augmented Generation (RAG)~\cite{10.5555/3495724.3496517} and deep research~\cite{10.1145/3655615,tongyideepresearchteam2025tongyideepresearchtechnicalreport,li-etal-2025-search}.
This view has motivated retrievers that can reason over complex queries~\cite{zhang2025qwen3embeddingadvancingtext,shao2025reasonir,chen2026AgentIR}, rather than relying on vanilla lexical retrievers such as BM25.
However, as frontier LLMs become increasingly capable of reasoning and tool use, these systems increasingly operate through an agentic loop, where LLMs receive feedback from their environments~\cite{yao2023react}.
Under this interaction pattern, retriever effectiveness may no longer be the only component worth optimizing at all costs.
We therefore revisit a central question in building such systems: \textit{Does a lexical retriever suffice as LLMs become more capable in an agentic loop?}
Without answering it, we again risk overemphasizing retriever improvement while overlooking other opportunities~\cite{yang2019critically}.

\begin{figure}[!t]
\centering
\begin{tikzpicture}
\begin{axis}[
    width=\columnwidth,
    height=0.72\columnwidth,
    set layers,
    axis on top=false,
    clip=false,
    xmode=log,
    xmin=20, xmax=1600,
    ymin=0, ymax=100,
    xlabel={Cost (\$)},
    ylabel={Accuracy (\%)},
    grid=major,
    major grid style={line width=.2pt, draw=gray!25},
    log basis x=10,
    xtick={25,50,100,200,400,800,1600},
    xticklabels={25,50,100,200,400,800,1600},
    ytick={10,20,30,40,50,60,70,80,90,100},
    extra x ticks={75,133.33,166.67,250,300,350,466.67,533.33,600,666.67,733.33,850,900,950,1080,1160,1240,1320,1400,1480,1560,1640,1720},
    extra x tick labels={},
    extra x tick style={
        tick style={draw=none},
        grid=major,
        major grid style={line width=.15pt, draw=gray!16}
    },
    tick label style={font=\fontsize{5}{6}\selectfont},  
    label style={font=\scriptsize},
]

\begin{pgfonlayer}{axis background}
  \begin{scope}[overlay]
    \clip (rel axis cs:0,0) rectangle (rel axis cs:1,1);

    \shade[
      shading=radial,
      inner color=green!18,
      outer color=white,
      opacity=0.7
    ]
    (rel axis cs:0.02,1.00) circle [radius=4cm];

    \shade[
      shading=radial,
      inner color=green!12,
      outer color=white,
      opacity=0.16
    ]
    (rel axis cs:0.20,0.84) circle [radius=3cm];

    \shade[
      shading=radial,
      inner color=green!8,
      outer color=white,
      opacity=0.12
    ]
    (rel axis cs:0.34,0.72) circle [radius=1cm];
  \end{scope}
\end{pgfonlayer}

\node[
  anchor=north west,
  text=green!50!black,
  font=\bfseries\fontsize{6}{7}\selectfont
] at (rel axis cs:0.04,0.95) {$\nwarrow$ better};

\addplot[
    only marks,
    mark=square*,
    mark size=1.5pt,
    color=blue!75,
    fill=blue!55,
]
coordinates {
    (28.92,68.07)  
    (55.08,71.43)  
};

\addplot[
    only marks,
    mark=square*,
    mark size=1.5pt,
    color={rgb,255:red,16; green,163; blue,127},
    fill={rgb,255:red,16; green,163; blue,127},
]
coordinates {
    (175.46,73.25)  
    (291.55,83.13)  
    (122.22,70.48)  
    (94.92,74.58)  
};

\addplot[
    only marks,
    mark=square*,
    mark size=1.5pt,
    color=orange!80!brown,
    fill=orange!60,
]
coordinates {
    (193.50,54.82)  
    (246.57,69.76)  
};

\addplot[
    only marks,
    mark=*,
    mark size=1.3pt,
    color=gray!70!black,
    fill=gray!55,
] coordinates {
    (836.35,50.84) 
    (740.79,66.27) 
    (400.36,58.31) 
    (360.71,73.01) 
    (138.64,19.04) 
    (99.92,28.67)  
    (1000,45.10)   
};

\addplot[
    only marks,
    mark=diamond*,
    mark size=2.0pt,
    color=gray!70!black,
    fill=gray!35,
] coordinates {
    (687.24,78.50) 
    (583.49,88.50) 
    (93.00,62.90) 
    (1016,80) 
};

\addplot[
    dashed,
    very thick,
    color=red!75!black,
] coordinates {
    (28.92,68.07)
    (55.08,71.43)
    (94.92,74.58)
    (291.55,83.13)
};

\addplot[
    only marks,
    mark=o,
    mark size=3.5pt,
    thin,
    color=red!75!black,
] coordinates {
    (28.92,68.07)
    (55.08,71.43)
    (94.92,74.58)
    (291.55,83.13)
};

\node[anchor=north, font=\fontsize{4}{5}\selectfont, align=center] at (axis cs:28.92,68.07)
{\code{ds-flash}};

\node[anchor=north, font=\fontsize{4}{5}\selectfont, align=center] at (axis cs:55.08,71.43)
{\code{ds-pro}};

\node[anchor=west,
      xshift=2pt,
      yshift=2pt,
      font=\fontsize{4}{5}\selectfont, 
      align=center
] at (axis cs:291.55,83.13)
{\code{gpt-5.5}};


\node[anchor=west,
      xshift=1pt,
      yshift=3pt,
      font=\fontsize{4}{5}\selectfont,
      align=left, 
      text=gray!40!black
] at (axis cs:173,73)
{\code{gpt-5.4}};

\node[anchor=west,
      xshift=0pt,
      yshift=-2pt,
      font=\fontsize{4}{5}\selectfont, 
      align=left, text=gray!40!black
] at (axis cs:122.22,70.48)
{\code{gpt-5.2}};

\node[anchor=south,
      xshift=0pt,
      yshift=0pt,
      font=\fontsize{4}{5}\selectfont, 
      align=left, text=gray!40!black
] at (axis cs:94.92,74.58)
{\code{gpt-5}};

\node[anchor=north,
      font=\fontsize{4}{5}\selectfont, 
      align=center, 
      text=gray!40!black
] at (axis cs:193.50,54.8)
{\code{haiku-4.5}};

\node[anchor=west,
      xshift=-6pt,
      yshift=-6pt,
      font=\fontsize{4}{5}\selectfont, 
      align=center, 
      text=gray!40!black
] at (axis cs:246.57,69.76)
{\code{opus-4.7}};

\node[
    anchor=north,
    xshift=6pt,
    yshift=0pt,
    font=\fontsize{3}{4}\selectfont,
    align=left,
    text=gray!40!black
] at (axis cs:1000,45.10)
{\code{gpt-5.2} w/ \code{qwen3}\\\citeauthor{meng2026revisiting}};

\node[
    anchor=south,
    xshift=6pt,
    yshift=0pt,
    font=\fontsize{3}{4}\selectfont,
    align=left,
    text=gray!40!black
] at (axis cs:687.24,78.50)
{\code{codex} w/ \code{bm25}\\\citeauthor{cao2026codingagentseffectivelongcontext}};

\node[
    anchor=south,
    xshift=4pt,
    yshift=0pt,
    font=\fontsize{3}{4}\selectfont,
    align=left,
    text=gray!40!black
] at (axis cs:583.49,88.50)
{\code{codex} w/ \code{cli}\\\citeauthor{cao2026codingagentseffectivelongcontext}};

\node[
    anchor=north,
    xshift=6pt,
    yshift=0pt,
    font=\fontsize{3}{4}\selectfont,
    align=left,
    text=gray!40!black
] at (axis cs:93,62.9)
{\code{DCI-lite}\\\citeauthor{li2026beyondsemantic}};

\node[
    anchor=south west,
    xshift=0pt,
    yshift=0pt,
    font=\fontsize{3}{4}\selectfont,
    align=left,
    text=gray!40!black
] at (axis cs:1016,80)
{\code{DCI-CC}\\\citeauthor{li2026beyondsemantic}};

\node[
    anchor=west,
    xshift=-1pt,
    yshift=6pt,
    font=\fontsize{3}{4}\selectfont,
    align=left,
    text=gray!40!black
] at (axis cs:836.35,50.84)
{\code{o3} w/ \code{bm25}\\\citeauthor{chen2025browsecompplusfairtransparentevaluation}};

\node[
    anchor=west,
    xshift=0pt,
    yshift=2pt,
    font=\fontsize{3}{4}\selectfont,
    align=left,
    text=gray!40!black
] at (axis cs:740.79,66.27)
{\code{o3} w/ \code{qwen3}\\\citeauthor{chen2025browsecompplusfairtransparentevaluation}};

\node[
    anchor=west, font=\fontsize{3}{4}\selectfont, align=left, text=gray!40!black
] at (axis cs:400.36,58.31)
{\code{gpt-5} w/ \code{bm25}\\\citeauthor{chen2025browsecompplusfairtransparentevaluation}};

\node[
    anchor=west,
    xshift=0pt,
    yshift=0pt, 
    font=\fontsize{3}{4}\selectfont, 
    align=left, 
    text=gray!40!black
] at (axis cs:360.71,73.01)
{\code{gpt-5} w/ \code{qwen3}\\\citeauthor{chen2025browsecompplusfairtransparentevaluation}};

\node[
    anchor=west, font=\fontsize{3}{4}\selectfont, align=left, text=gray!40!black
] at (axis cs:138.64,19.04)
{\code{gemini-2.5-pro} w/ \code{bm25}\\\citeauthor{chen2025browsecompplusfairtransparentevaluation}};

\node[
    anchor=west, font=\fontsize{3}{4}\selectfont, align=left, text=gray!40!black
] at (axis cs:99.92,28.67)
{\code{gemini-2.5-pro} w/ \code{qwen3}\\\citeauthor{chen2025browsecompplusfairtransparentevaluation}};

\end{axis}

\node[
    anchor=north,
    yshift=0mm,
    align=center,
    inner sep=0pt,
] at (current bounding box.south) {%
\fontsize{5}{6}\selectfont
\begin{tabular}{@{}c@{\hspace{0.35em}}l@{\hspace{1.2em}}c@{\hspace{0.35em}}l@{\hspace{1.2em}}c@{\hspace{0.35em}}l@{}}
\tikz{\filldraw[draw=blue!75,fill=blue!55] (0,0) rectangle +(0.11,0.11);} & \textsc{Pi-Serini} (DeepSeek) &
\tikz{\filldraw[draw={rgb,255:red,16; green,163; blue,127},fill={rgb,255:red,16; green,163; blue,127}] (0,0) rectangle +(0.11,0.11);} & \textsc{Pi-Serini} (OpenAI) &
\tikz{\filldraw[draw=orange!80!brown,fill=orange!60] (0,0) rectangle +(0.11,0.11);} & \textsc{Pi-Serini} (Anthropic) \\
\multicolumn{6}{c}{} \\
\tikz{\filldraw[draw=gray!70!black,fill=gray!55] (0.055,0.055) circle (0.055);} & Prior work &
\tikz{\filldraw[draw=gray!70!black,fill=gray!35] (0.055,0.055) -- (0.11,0.11) -- (0.165,0.055) -- (0.11,0) -- cycle;} & Coding agent (filesystem) &
\tikz{\draw[red!75!black,dashed,very thick] (0,0.055) -- +(0.22,0);} & Pareto frontier \\
\end{tabular}%
};
\end{tikzpicture}
\caption{Accuracy vs. Cost trade-off on BrowseComp-Plus. 
\code{ds} abbreviates \code{deepseek-v4} model, and \code{qwen3} abbreviates \code{qwen3-embed-8b} retriever.
All \textsc{Pi-Serini} systems are paired with BM25.\protect\footnotemark
}
\label{fig:accuracy-cost-tradeoff-loggrid}
\vspace{-0.5cm}
\end{figure}
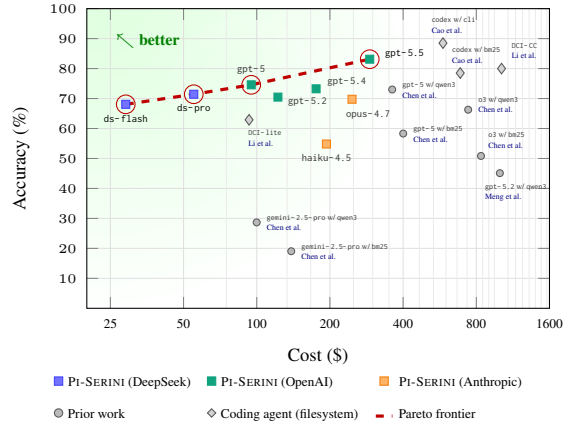
\footnotetext{Results from~\citet{cao2026codingagentseffectivelongcontext} are based on sampled evaluations and provided for reference. See Section~\ref{retrieve_or_not}.}

We begin by testing whether previous lexical baselines were merely ill-configured or whether they are genuinely incapable of finding relevant documents. 
We verify the BM25 parameter settings and the retrieval depth to increase the likelihood that relevant documents remain in retrieved results, rather than being discarded before they are returned to and processed by the LLM.

However, because LLMs are often constrained by limited context windows, they need an interface that helps them extract useful information for subsequent actions.
We therefore introduce \textsc{Pi-Serini}, a deliberately minimal search agent.
The agent is equipped with three search-related tools---\code{search}, \code{read\_search\_results}, and \code{read\_document}---so that document retrieval, result browsing, and document reading become distinct decisions. 
By leveraging the tool-use abilities of LLMs, \textsc{Pi-Serini} uses this tool interface to cache a retrieved ranking, explore deeper retrieval results, and decide which evidence enters its context window.

In addition, deep research is inherently time-sensitive, requiring systems to balance answer quality against realistic latency and cost constraints.
To better match user-facing deep research workloads, we use time-budget steering instead of the fixed iteration cap used in prior work~\cite{chen2025browsecompplusfairtransparentevaluation,meng2026revisiting}, allowing the agent to terminate under a hard wall-clock time budget.

On BrowseComp-Plus~\cite{chen2025browsecompplusfairtransparentevaluation}, under time-budget steering, \textsc{Pi-Serini} with \code{gpt-5.5} achieves 83.1\% answer accuracy and 94.7\% surfaced evidence recall, outperforming released search agents based on dense retrievers.
Ablations show that BM25 tuning improves answer accuracy by 18.0\% and surfaced evidence recall by 11.1\% over the default BM25 setting, while increasing retrieval depth further improves surfaced evidence recall by 25.3\% over the shallow-retrieval setting.

Finally, \textsc{Pi-Serini} benefits from a prefix-cache-friendly agentic loop, allowing many repeated input tokens to be served from cache and making prefix caching central to its cost efficiency.
BM25 tuning and wall-clock time budgeting provide additional cost savings.
Figure~\ref{fig:accuracy-cost-tradeoff-loggrid} summarizes the resulting accuracy--cost trade-off, showing that \textsc{Pi-Serini} consistently lies in the efficient region.

Overall, the BrowseComp-Plus experiments suggest that, with more capable LLMs in an agentic loop, BM25 can be sufficient for effective deep research when it is well-configured, used at sufficient retrieval depth, and paired with a tool interface that helps the agent manage cached retrieval ranking.
Our contributions are as follows:
\begin{itemize}[
  leftmargin=*,
  topsep=0.25em,
  itemsep=0.15em,
  parsep=0pt,
  partopsep=0pt
]
    \item We introduce \textsc{Pi-Serini}, a search agent that manages cached rankings for deeper retrieval and selective evidence acquisition.
    \item We reassess BM25 on BrowseComp-Plus by separating lexical retrieval capacity from configuration and retrieval depth, showing that both greatly affect answer accuracy and evidence recall.
    \item We show that, on BrowseComp-Plus, BM25-based search agents can match or outperform dense-retriever agents, while reducing evaluation cost by 3.3$\times$--10$\times$.
\end{itemize}

\section{Related Work}

\paragraph{Deep Research.}
Retrieval-Augmented Generation (RAG)~\cite{10.5555/3495724.3496517} grounds answer generation in external documents, but complex information-seeking tasks often require more than a single retrieval step.
They involve iterative search, document inspection, and evidence synthesis across multiple sources, motivating search agents that extend RAG into multi-step retrieval-and-reasoning processes~\cite{shi2025deepresearchsystematicsurvey,tongyideepresearchteam2025tongyideepresearchtechnicalreport,asai2024selfrag}.
Recent work has improved these agents through reasoning enhancement~\cite{10.1145/3774904.3792235,wen2026reinforcement,shao2024deepseekmathpushinglimitsmathematical} and tool-use optimization~\cite{wu2024avatar,qian2026toolrl}. 
As these systems become more capable, their evaluation has also shifted toward more realistic research settings:
Benchmarks such as BrowseComp~\cite{wei2025browsecompsimplechallengingbenchmark} and BrowseComp-ZH~\cite{zhou2025browsecompzhbenchmarkingwebbrowsing} evaluate open-web browsing ability, while BrowseComp-Plus provides a fixed-corpus setting for controlled comparison of search agents~\cite{chen2025browsecompplusfairtransparentevaluation}.
These benchmarks make retrieval and evidence interaction central to agent performance, raising the question of how much observed performance depends on the retrieval model itself.

\paragraph{Retriever for Deep Research.}
Search agents often use dense retrievers, sparse retrievers, and rerankers to improve early-stage ranking accuracy~\cite{zhang2025qwen3embeddingadvancingtext}.
Recent reasoning-aware retrievers further target complex queries by capturing implicit intent and resolving multi-hop information needs~\cite{shao2025reasonir,chen2026AgentIR}.
By contrast, lexical methods such as BM25 are commonly used as non-dense baselines, and their weaker end-to-end performance is often attributed to semantic mismatch.

Related analyses have examined retrieval components in search agents: SAGE~\cite{hu2026sagebenchmarkingimprovingretrieval} studies retrieval effectiveness with LLM-based retrievers, while other work~\cite{sharifymoghaddam2026rerankreasonanalyzingreranking,meng2026revisiting} studies how reranking and document ordering affect search agents in deep research.
However, these studies do not isolate whether BM25's apparent weakness comes from lexical retrieval itself or surrounding design choices such as configuration and retrieval depth.
Our work addresses this gap by separating BM25 capacity from these surrounding choices.
We therefore revisit the question: \textit{Does a lexical retriever suffice as LLMs become more capable in an agentic loop?}

\section{\textsc{Pi-Serini}}

\textsc{Pi-Serini} is a search agent for isolating agent--retriever interaction.
The LLM agent runs a ReAct loop, while a retrieval controller mediates all access to an \textsc{Anserini} BM25 backend.
This controller is the main isolation point: it exposes a constrained tool API, maintains cached ranking from retrieval, and controls how retrieved evidence enters the agent context.
Appendix~\ref{app:system-setup} gives the exact prompt, backend configuration, cache behavior, and runtime policy used in our experiments.

\subsection{Preliminary}
We formulate deep research as an iterative retrieval-and-reasoning problem in a ReAct-style agentic loop~\cite{yao2023react}.
Given a user query, a search agent, commonly materialized by an LLM, iteratively reasons, calls tools, observes their outputs, and eventually returns a final response.

The agent operates over an interaction trajectory:
\[
\mathcal{H}_T = (\tau_1, a_1, o_1, \ldots, \tau_t, a_t, o_t, \ldots, \tau_T, a_T)
\]
where $\tau_t$ is the reasoning trace generated at turn $t$, $a_t$ is the action selected by the agent, and $o_t$ is the observation returned by the environment.
At each turn, the agent policy $\pi$ samples a reasoning trace and an action conditioned on the previous interaction history:
\[
\tau_t, a_t \sim \pi(\cdot \mid \mathcal{H}_{t-1})
\]

For intermediate turns $(t < T)$, the action $a_t$ is a tool call or reasoning, and the environment returns an observation $o_t$.
The final action $a_T$ produces the response and terminates the loop.
We treat the outputs as both the final response $R$ and the set of documents $D$ observed during the retrieval process.

\subsection{Components}

\textsc{Pi-Serini} has three design elements.
First, it separates retrieval, result browsing, and document reading into distinct tool calls.
Second, it records the agent's interaction trajectory at multiple evidence-access levels.
Third, it uses time-budget steering to ensure that the agent answers before timeout.

\paragraph{Retrieval Controller.}
Naive search agents often use shallow retrieval settings that insert the full texts of all ranked documents directly into the context window.
This removes the need for explicit evidence selection, but conflates retrieval depth with context management and makes deeper retrieval impractical.
In contrast, \textsc{Pi-Serini} leverages LLM tool use through separate tools for retrieval, browsing, and reading, allowing the agent to cache ranked documents and selectively control which evidence enters its context window.
Our tool interface follows the workflow in Appendix~\ref{app:retrieval-tools}: search first, browse the cached ranking when useful, and selectively read promising documents.
The action $a$ can be one of three search-related tool calls:
\begin{itemize}[
  leftmargin=*,
  topsep=0.25em,
  itemsep=0.15em,
  parsep=0pt,
  partopsep=0pt
]
  \item \code{search}: issues a lexical query, retrieves up to 1000 ranked documents, caches the ranking under a session-local \code{search\_id}, and exposes only the top 5 excerpts on the initial call;
  \item \code{read\_search\_results}: browses a cached ranking by \code{search\_id} using rank pagination, allowing the agent to inspect results without issuing a new backend query;
  \item \code{read\_document}: fetches one document by identifier using line-based pagination, allowing the agent to read only the relevant text pieces.
\end{itemize}

\label{piserini:logging}
\paragraph{Trajectory Logging.}
Each run produces a structured per-query artifact that records the full tool call and reasoning trajectory.
Rather than tracking only a single document set returned by the search tool, \textsc{Pi-Serini} records four document sets:
\begin{itemize}[
  leftmargin=*,
  topsep=0.25em,
  itemsep=0.15em,
  parsep=0pt,
  partopsep=0pt
]
  \item $D_{\text{surfaced}}$: documents returned by \code{search}.
  \item $D_{\text{previewed}}$: documents whose excerpts were shown through \code{read\_search\_results}.
  \item $D_{\text{opened}}$: documents read via \code{read\_document}.
  \item $D_{\text{cited}}$: documents cited in the final answer.
\end{itemize}
These tiers distinguish what the system made available, what the agent chose to inspect, and what it ultimately used.

\paragraph{Time-Budget Steering.}
We impose a timeout rather than a fixed iteration cap on the agentic loop.
Each query run is subject to a timeout of $T$ seconds.

To encourage completion before hard termination, \textsc{Pi-Serini} uses a two-stage time-budget policy.
At $0.7T$, the system injects a submission steer that instructs the agent to stop using tools and produce its best answer from the evidence collected so far.
From that point onward, all three tools are blocked.
If the agent still does not submit by time $T$, the process is terminated and the query is marked as timed out.

\section{Experimental Setup}
We study our research question---\textit{Does a lexical retriever suffice as LLMs become more capable in an agentic loop?}---by evaluating search agents that pair different LLMs with either dense retrievers or BM25, measuring both retrieval effectiveness and final answer quality.

\subsection{Evaluation Settings}

\paragraph{Dataset.}
We evaluate \textsc{Pi-Serini} on the widely used BrowseComp-Plus~\cite{chen2025browsecompplusfairtransparentevaluation}, a deep research benchmark consisting of 830 queries and 100,195 documents. 
On average, each query is associated with 6.1 evidence documents and 2.9 gold documents. 
Each document contains an average of 5,179.2 words and 32,296.2 characters.

Evidence documents are documents required to answer the query, while gold documents are a stricter subset that both support answering and semantically contain the final answer.

\paragraph{Baselines.}
We compare \textsc{Pi-Serini} with several representative BrowseComp-Plus baseline systems. 
The released systems pair \code{o3} and \code{gpt-5} with either \code{qwen3-embed-8b}~\cite{zhang2025qwen3embeddingadvancingtext} or BM25.
We also report the baseline from the work of~\citet{meng2026revisiting}, which also uses \code{qwen3-embed-8b} retriever but replaces the reasoning model with \code{gpt-5.2}. 
Finally, we report the numbers of AgentIR~\cite{chen2026AgentIR}, a reasoning-intensive retriever trained for deep research.

Unlike \textsc{Pi-Serini}, which exposes separate \code{search}, \code{read\_search\_results}, and \code{read\_document} tools, the baseline agents use a single retriever tool that directly returns the top-$k$ search results, with $k=5$.
Under our logging protocol, we therefore only consider $D_{\text{surfaced}}$ for baseline agents, since they do not use additional tools beyond \code{search} unless specified otherwise.

\paragraph{Metrics.}

Following BrowseComp-Plus's evaluation settings, we report both answer-quality metrics and retrieval-behavior metrics.
For answer evaluation, we use an LLM judge.
For each query, the judge receives the question, the agent's final response, and the benchmark-provided correct answer, then determines whether the extracted final answer is semantically equivalent to the correct answer.
The judge is run with \code{gpt-5.3-codex}; Appendix~\ref{app:gold-answer-judge} gives the exact configuration and prompt.
Specifically, we report:

\begin{itemize}[
  leftmargin=*,
  topsep=0.25em,
  itemsep=0.15em,
  parsep=0pt,
  partopsep=0pt
]
    \item Accuracy: the fraction of queries whose final answers are judged correct by the LLM judge;

    \item Calibration Error: the discrepancy between the model's confidence and empirical correctness.
\end{itemize}

\smallskip\noindent
For retrieval evaluation, we report recall over the logged document sets introduced in Section~\ref{piserini:logging}:

\begin{itemize}[
  leftmargin=*,
  topsep=0.25em,
  itemsep=0.15em,
  parsep=0pt,
  partopsep=0pt
]
    \item Surfaced Recall: recall computed over the document set $D_{\text{surfaced}}$;

    \item Previewed Recall: recall computed over the document set $D_{\text{previewed}}$;

    \item Behavior Recall: recall computed over the union of the document sets $D_{\text{opened}} \cup D_{\text{cited}}$.
\end{itemize}

\subsection{\textsc{Pi-Serini} Implementation Details}

\paragraph{Retriever.}

Documents are indexed using BM25 via \textsc{Anserini}~\cite{10.1145/3239571} over the BrowseComp-Plus corpus. 
We use tuned BM25 parameters with $k_1 = 25$ and $b = 1$, which are selected to better support long-document retrieval in a deep research setting. The BM25 retrieval depth is set to 1000. 
This retrieval configuration is held fixed across all \textsc{Pi-Serini} runs.
Further details on BM25 tuning are provided in Section~\ref{abl:tuning-bm25}.

\paragraph{Models.}

We evaluate \textsc{Pi-Serini} using a diverse set of frontier LLMs with better reasoning and tool-use abilities, reflecting the class of sufficiently capable LLMs considered in our research question.
For DeepSeek, we use \code{deepseek-v4-flash} and \code{deepseek-v4-pro}. 
For Anthropic, we use \code{claude-haiku-4.5} and \code{claude-opus-4.7}. 
For OpenAI, we use \code{gpt-5}, \code{gpt-5.2}, \code{gpt-5.4}, \code{gpt-5.4-mini}, and \code{gpt-5.5}. 
This model set across different providers allows us to evaluate whether \textsc{Pi-Serini}'s design remains effective across different LLM families and scales.

\paragraph{Agent Harness.}
\label{harness}
We build \textsc{Pi-Serini} on top of \textsc{Pi}~\cite{pi_mono_2025}, a minimal but malleable agent harness without sub-agents or additional orchestration modules.
We repurpose \textsc{Pi} as our search agent by removing its system prompts and built-in tools, then design our own.
We configure it with our deep research prompt that instructs it to issue queries, inspect retrieval results through our tools, and synthesize an answer.
Per-query timeout is set to $T = 300$ seconds.

\begin{table*}[t]
\centering
\small

\begin{threeparttable} 
\setlength{\tabcolsep}{4.2pt}
\renewcommand{\arraystretch}{1.12}

\resizebox{0.9\linewidth}{!}{%
\begin{tabular}{
@{}c
l
l
@{\hspace{0.8em}}rrr
@{\hspace{1.0em}}rr
@{\hspace{1.0em}}rr
@{\hspace{1.0em}}rr
@{}
}
\toprule
\multirow{2}{*}{}
& \multirow{2}{*}{\textbf{LLM}}
& \multirow{2}{*}{\textbf{Retriever}}
& \multicolumn{3}{c}{\textbf{Answer Quality}}
& \multicolumn{2}{c}{\makecell[c]{\textbf{Surfaced}\\\textbf{Recall}}}
& \multicolumn{2}{c}{\makecell[c]{\textbf{Previewed}\\\textbf{Recall}}}
& \multicolumn{2}{c}{\makecell[c]{\textbf{Behavior}\\\textbf{Recall}}} \\
\cmidrule(lr){4-6}
\cmidrule(lr){7-8}
\cmidrule(lr){9-10}
\cmidrule(l){11-12}
& & &
\textbf{Acc.}
& \textbf{Calib.}
& \textbf{Cost (\$)}
& \textbf{Evi.}
& \textbf{Gold}
& \textbf{Evi.}
& \textbf{Gold}
& \textbf{Evi.}
& \textbf{Gold} \\
\midrule

\multicolumn{12}{@{}l}{\citet{chen2025browsecompplusfairtransparentevaluation}$^{\dagger}$} \\
\addlinespace[2pt]

\tabrowmark{main:o3-bm25}{a}
& \code{o3}
& \code{bm25}
& 50.8 & 39.1 & 836.4
& 56.6 & 61.7
& - & -
& - & - \\

\tabrowmark{main:o3-qwen3}{b}
& \code{o3}
& \code{qwen3-embed-8b}
& 66.3 & 32.7 & 740.8
& 73.2 & 76.3
& - & -
& - & - \\

\tabrowmark{main:gpt5-bm25}{c}
& \code{gpt-5}
& \code{bm25}
& 58.3 & 13.5 & 400.4
& 61.7 & 66.5
& - & -
& - & - \\

\tabrowmark{main:gpt5-qwen3}{d}
& \code{gpt-5}
& \code{qwen3-embed-8b}
& 73.0 & 9.7 & 360.7
& 79.0 & \textbf{81.3}
& - & -
& - & - \\

\midrule
\multicolumn{12}{@{}l}{\citet{meng2026revisiting}$^{\ddagger}$} \\
\addlinespace[2pt]

\tabrowmark{main:gpt52-qwen3}{e}
& \code{gpt-5.2}
& \code{qwen3-embed-8b}
& 45.1 & - & $1 \sim 2$k
& - & 74.7
& - & -
& - & - \\

\midrule
\multicolumn{12}{@{}l}{\citet{chen2026AgentIR}$^{\ddagger}$} \\
\addlinespace[2pt]

\tabrowmark{main:tongyi-agentir}{f}
& \code{Tongyi-DR}
& \code{AgentIR-4B}
& 68.1 & - & -
& \textbf{79.2} & -
& - & -
& - & - \\

\midrule
\multicolumn{12}{@{}l}{\textsc{Pi-Serini} (ours)} \\
\addlinespace[2pt]

\tabrowmark{main:claude-haiku-bm25}{g}
& \code{claude-haiku-4.5}
& \multirow{8}{*}{\code{bm25}}
& 54.8 & 17.3 & 193.5
& 94.1 & \textbf{95.4}
& 58.2 & 60.1
& 40.9 & 46.3 \\

\tabrowmark{main:claude-opus-bm25}{h}
& \code{claude-opus-4.7}
& 
& 69.8 & 10.2 & 246.6
& 81.2 & 86.8
& 43.3 & 52.5
& 30.4 & 43.0 \\

\tabrowmark{main:pi-gpt5-bm25}{i}
& \code{gpt-5}
&
& 74.6 & 7.2 & 94.9
& 90.5 & 93.8
& 62.7 & 70.0
& 45.7 & 56.2 \\

\tabrowmark{main:gpt52-bm25}{j}
& \code{gpt-5.2}
&
& 70.5 & 6.2 & 122.2
& 89.9 & 92.8
& 60.5 & 67.3
& 44.8 & 54.5 \\

\tabrowmark{main:gpt54-mini-bm25}{k}
& \code{gpt-5.4-mini}
&
& 68.1 & 13.7 & 86.2
& 91.9 & 94.1
& 60.1 & 65.4
& 43.1 & 52.1 \\

\tabrowmark{main:gpt54-bm25}{l}
& \code{gpt-5.4}
&
& 73.3 & 9.2 & 175.5
& 93.8 & 95.3
& 70.3 & 66.9
& 51.8 & 58.1 \\

\tabrowmark{main:gpt55-bm25}{m}
& \code{gpt-5.5}
&
& \textbf{83.1} & 15.7 & 291.6
& \textbf{94.7} & 94.4
& 73.6 & 72.9
& 58.9 & 56.1 \\

\tabrowmark{main:deepseek-flash-bm25}{n}
& \code{deepseek-v4-flash}
&
& 68.1 & 15.5 & \textbf{28.9}
& 94.5 & 95.7
& 67.9 & 69.9
& 55.2 & 60.6 \\

\tabrowmark{main:deepseek-pro-bm25}{o}
& \code{deepseek-v4-pro}
&
& 71.4 & 7.0 & 55.1
& 91.3 & 92.5
& 60.0 & 63.0
& 45.4 & 50.6 \\

\bottomrule
\end{tabular}%
}

\caption{Answer quality, retrieval behavior, and cost on BrowseComp-Plus.}
\label{tab:main-results}

\begin{tablenotes}
\footnotesize
\item[$\dagger$] Results from \citet{chen2025browsecompplusfairtransparentevaluation} are obtained by evaluating their released runs using our own judge.
\item[$\ddagger$] Results from \citet{meng2026revisiting} and \citet{chen2026AgentIR} are taken from the reported numbers in the original papers.
\end{tablenotes}

\end{threeparttable}
\end{table*}

\section{Main Experimental Results}
\paragraph{Answer Quality.}

Table~\ref{tab:main-results} reports the answer accuracy of all systems.
We first compare systems that use the same LLM, \code{gpt-5}, with BM25.
Under this setting, \textsc{Pi-Serini} improves answer accuracy from 58.3\% for the released \code{gpt-5} + \code{BM25} baseline in row \tabrow{main:gpt5-bm25} to 74.6\% in row \tabrow{main:pi-gpt5-bm25}.
It also slightly exceeds the released \code{gpt-5} + \code{qwen3-embed-8b} dense-retriever baseline in row \tabrow{main:gpt5-qwen3}.
This comparison suggests that previously released baselines understate the potential of BM25, and that improving its configuration and increasing retrieval depth makes it a competitive alternative to dense retrievers for deep research.

Among BM25-based systems, \textsc{Pi-Serini} with \code{gpt-5.5} in row \tabrow{main:gpt55-bm25} achieves the highest accuracy, at 83.13\%.
This is also the highest accuracy in Table~\ref{tab:main-results}, further showing that a BM25-based search agent can achieve highly competitive performance when BM25 is well-configured, used with sufficient retrieval depth, and paired with more capable LLMs in an agentic loop.

On the other hand, using a dense retriever does not necessarily translate into better answer quality. 
The system from \citet{meng2026revisiting}, which pairs \code{gpt-5.2} with \code{qwen3-embed-8b} in row \tabrow{main:gpt52-qwen3}, achieves only 45.1\% accuracy. 
In contrast, \textsc{Pi-Serini} with the same LLM, \code{gpt-5.2}, reaches 70.5\% accuracy using BM25 in row \tabrow{main:gpt52-bm25}. 
This gap suggests that final answer quality is not determined by retriever choice alone, but also by how the agent searches, manages context, and interacts with retrieved documents. 
The same pattern is also observed for the AgentIR-4B baseline in row \tabrow{main:tongyi-agentir}.

Overall, these results address our research question in terms of final answer quality, showing that BM25 can achieve competitive answer accuracy in deep research when its parameters are well-configured, its retrieval depth is sufficient, and it is paired with more capable LLMs in an agentic loop.

We make two additional observations from the answer-quality results. First, the DeepSeek-based agents provide a strong cost--performance tradeoff: \code{deepseek-v4-flash} achieves 68.1\% accuracy with only \$28.9 total cost in row \tabrow{main:deepseek-flash-bm25}. Second, there is a substantial gap between \code{claude-opus-4.7} and \code{gpt-5.5}, despite both being high-cost frontier models. \code{claude-opus-4.7} achieves 69.8\% accuracy in row \tabrow{main:claude-opus-bm25}, while \code{gpt-5.5} reaches 83.1\% in row \tabrow{main:gpt55-bm25}. We further analyze their behavioral differences in Section~\ref{dis:search_behavior}.

\paragraph{Retrieval Behavior.}
Table~\ref{tab:main-results} compares retrieval behavior across systems.
For surfaced recall, all \textsc{Pi-Serini} variants substantially outperform the released baselines, with most runs achieving over 90\% recall on both evidence and gold documents. 
Notably, \code{gpt-5.5} + \code{BM25} in row \tabrow{main:gpt55-bm25} reaches 94.7 and 94.4 on the evidence and gold documents, respectively, compared to 79.21 and 81.34 for the dense-retriever baselines of rows \tabrow{main:gpt5-qwen3} and \tabrow{main:tongyi-agentir}.

Under the same \code{gpt-5} + \code{BM25} setting, \textsc{Pi-Serini} also exposes more evidence to the agent.
Compared with the released baseline in row \tabrow{main:gpt5-bm25}, \textsc{Pi-Serini} in row \tabrow{main:pi-gpt5-bm25} improves previewed recall of gold documents from 66.5 to 70.0, while maintaining similar previewed evidence-document recall.
With \code{gpt-5.5}, \tabrow{main:gpt55-bm25} further achieves the highest previewed recall among BM25-based systems, reaching 73.6 for evidence documents and 72.9 for gold documents.
Because agents can often synthesize answers from previewed excerpts without calling \code{read\_document}, higher previewed recall need not yield proportional gains in behavior recall.

Overall, these results answer our research question from the perspective of retrieval effectiveness: a lexical retriever can retrieve the evidence needed for deep research when paired with more capable LLMs in an agentic loop, provided that BM25 is well-configured and used with sufficient retrieval depth.
At the same time, the gap between surfaced and previewed recall shows that the remaining challenge lies in helping the agent navigate cached ranking and allocate context to the right evidence.

\paragraph{Cost Analysis.}
Prior work on BrowseComp-Plus has shown that deep research evaluation can be expensive~\cite{chen2025browsecompplusfairtransparentevaluation,meng2026revisiting}. 
\citet{meng2026revisiting}, for example, report that one \code{gpt-5.2} experiment in row \tabrow{main:gpt52-qwen3} costs roughly \$1000--\$2000 USD because many queries require many iterations for an agent to reach its conclusion.

\textsc{Pi-Serini} substantially lowers this cost while preserving competitive performance.
Under the same \code{gpt-5} + \code{BM25} setting, \textsc{Pi-Serini} reduces the cost from \$400.4 in the released baseline row \tabrow{main:gpt5-bm25} to \$94.9 in row \tabrow{main:pi-gpt5-bm25}.
Even when paired with more expensive LLMs, \textsc{Pi-Serini} remains cost-efficient.
For example, \code{gpt-5.5} + \code{BM25} in row \tabrow{main:gpt55-bm25} achieves the highest accuracy in the table at a cost of \$291.6, which is still lower than the released \code{gpt-5} baselines in rows \tabrow{main:gpt5-bm25} and \tabrow{main:gpt5-qwen3}.
Similarly, \textsc{Pi-Serini} with \code{gpt-5.2} costs \$122.2, far below the roughly \$1000--\$2000 reported for the \code{gpt-5.2} experiment in row \tabrow{main:gpt52-qwen3}.

These cost reductions matter because deep research studies depend on repeated evaluation. 
By lowering the cost of each full benchmark run, \textsc{Pi-Serini} makes ablations and future studies of deep research more practical.
We further analyze the factors contributing to the cost savings in Section~\ref{abl:cost-efficiency}.

\begin{table}[t]
\centering
\small
\setlength{\tabcolsep}{5pt}
\resizebox{\linewidth}{!}{
\begin{tabular}{lrrrr}
\toprule
\textbf{System} & \textbf{Total} & \textbf{Search} & \textbf{Read} & \textbf{Browse} \\
\midrule

\multicolumn{5}{l}{\citet{chen2025browsecompplusfairtransparentevaluation}} \\
\addlinespace[2pt]
\code{o3} + \code{BM25} & 25.9 & 25.9 & -- & -- \\
\code{o3} + \code{qwen3-embed-8b} & 24.0 & 24.0 & -- & -- \\
\code{gpt-5} + \code{BM25} & 23.2 & 23.2 & -- & -- \\
\code{gpt-5} + \code{qwen3-embed-8b} & 21.7 & 21.7 & -- & -- \\

\midrule

\multicolumn{5}{l}{\citet{chen2026AgentIR}} \\
\addlinespace[2pt]
\code{Tongyi-DR} + \code{AgentIR-4B} & 27.9 & 24.5 & 3.4 & -- \\

\midrule

\multicolumn{5}{l}{\citet{meng2026revisiting}} \\
\addlinespace[2pt]
\code{gpt-5.2} + \code{qwen3-embed-8b} & 73.8 & 73.8 & -- & -- \\

\midrule

\multicolumn{5}{l}{\textsc{Pi-Serini (ours)}} \\
\addlinespace[2pt]
\code{claude-haiku-4.5} + \code{BM25} & 41.3 & 31.8 & 8.8 & 0.6 \\
\code{claude-opus-4.7} + \code{BM25} & 9.0 & 6.8 & 2.1 & 0.1 \\
\code{gpt-5} + \code{BM25} & 15.2 & 11.1 & 3.9 & 0.2  \\
\code{gpt-5.2} + \code{BM25} & 17.2 & 12.6 & 3.9 & 0.7  \\
\code{gpt-5.4-mini} + \code{BM25} & 24.7 & 20.2 & 4.2 & 0.3 \\
\code{gpt-5.4} + \code{BM25} & 23.1 & 17.9 & 4.3 & 0.9 \\
\code{gpt-5.5} + \code{BM25} & 19.3 & 13.5 & 5.0 & 0.8 \\
\code{deepseek-v4-flash} + \code{BM25} & 33.2 & 26.1 & 7.0 & 0.1 \\
\code{deepseek-v4-pro} + \code{BM25} & 18.6 & 14.2 & 4.3 & 0.1 \\

\bottomrule
\end{tabular}
}
\caption{Average tool calls per query for all systems}
\label{tab:bm25-tool-calls}
\end{table}

\paragraph{Tool Usage.}
Table~\ref{tab:bm25-tool-calls} further compares how different search agents allocate their tool budget.
The results show that, except for the \code{claude-haiku-4.5}-based agent, most \textsc{Pi-Serini} systems use a similar or smaller number of total tool calls compared to the BrowseComp-Plus released baselines. 
This pattern is especially clear under the same \code{gpt-5} + \code{BM25} setting.
Compared with the released baseline, which uses 23.2 tool calls per query, \textsc{Pi-Serini} uses only 15.2. 
In contrast, the \code{gpt-5.2} + \code{qwen3-embed-8b} system reported by \citet{meng2026revisiting} uses a dense retriever but requires a substantially more tool calls, 73.8 per query, while not achieving better answer accuracy than \textsc{Pi-Serini}. 
This indicates that replacing dense retrievers with BM25 does not necessarily increase the number of tool calls, while still maintaining competitive retrieval effectiveness and answer accuracy, as shown in Table~\ref{tab:main-results}.

\section{Ablation Study}

\paragraph{Retrieval Depth.}

\begin{figure}[t]
\centering

\begin{tikzpicture}
\begin{axis}[
    width=0.9\linewidth,
    height=5cm,
    xlabel={Top-$k$},
    ylabel={Recall (\%)},
    xmode=log,
    log basis x=10,
    xmin=5, xmax=1000,
    ymin=65, ymax=100,
    xtick={5,10,20,50,100,500,1000},
    xticklabels={5,10,20,50,100,500,1000},
    xlabel style={font=\footnotesize},
    ylabel style={font=\footnotesize},
    legend style={
        at={(0.5,-0.3)},
        anchor=north,
        legend columns=2,
        font=\scriptsize,
        tick label style={font=\scriptsize},
        legend style={font=\scriptsize},
    },
    grid=major,
    thick
]

\addplot[
    blue,
    mark=o
] coordinates {
    (5,70.48)
    (10,76.84)
    (20,82.70)
    (50,86.16)
    (100,86.22)
    (500,94.16)
    (1000,95.78)
};
\addlegendentry{Surfaced (Evi.)}

\addplot[
    blue,
    mark=square
] coordinates {
    (5,73.12)
    (10,81.82)
    (20,87.93)
    (50,87.67)
    (100,90.45)
    (500,96.83)
    (1000,97.33)
};
\addlegendentry{Surfaced (Gold)}

\addplot[
    red,
    mark=o
] coordinates {
    (5,70.48)
    (10,72.32)
    (20,74.52)
    (50,74.67)
    (100,74.07)
    (500,70.60)
    (1000,70.89)
};
\addlegendentry{Previewed (Evi.)}

\addplot[
    red,
    mark=square
] coordinates {
    (5,73.12)
    (10,77.33)
    (20,79.73)
    (50,76.49)
    (100,79.82)
    (500,71.45)
    (1000,74.78)
};
\addlegendentry{Previewed (Gold)}

\end{axis}
\end{tikzpicture}

\caption{Effect of the number of documents ($k$) returned by the \code{search} tool in \textsc{Pi-Serini} with \code{gpt-5.4}}
\label{fig:ablation-search-k}
\end{figure}
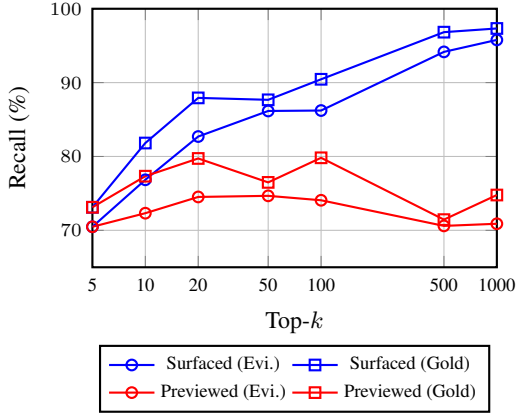

We vary the number of documents $k$ returned by \code{search} to understand how much of the recall gain comes from increased retrieval depth.

Figure~\ref{fig:ablation-search-k} shows that increasing $k$ from 5 to 100 substantially improves surfaced recall from 70.5\% to 86.22\%.
Previewed recall also improves over the same range, from 70.5\% to 74.1\%.
When $k$ is further increased to 1000, surfaced recall continues to rise, reaching 95.8\%.
In contrast, previewed recall does not continue to improve with larger $k$ and reaches its highest value of 74.7\% at $k=50$.

These results show that retrieval depth is a major driver of surfaced recall in \textsc{Pi-Serini}.
Increasing $k$ gives BM25 enough first-stage capacity to place relevant evidence documents in the cached ranking, but the saturation of previewed recall suggests that larger cached rankings do not automatically lead the agent to inspect more relevant evidence.

\paragraph{Tuning BM25 for Long Documents.}
\label{abl:tuning-bm25}
BM25 defaults are usually chosen for shorter documents, but BrowseComp-Plus documents are long and noisy.
We therefore test whether \textsc{Pi-Serini}'s gains depend on retuning BM25 for this setting.

\begin{figure}[t]
\centering
\includegraphics[width=0.9\linewidth]{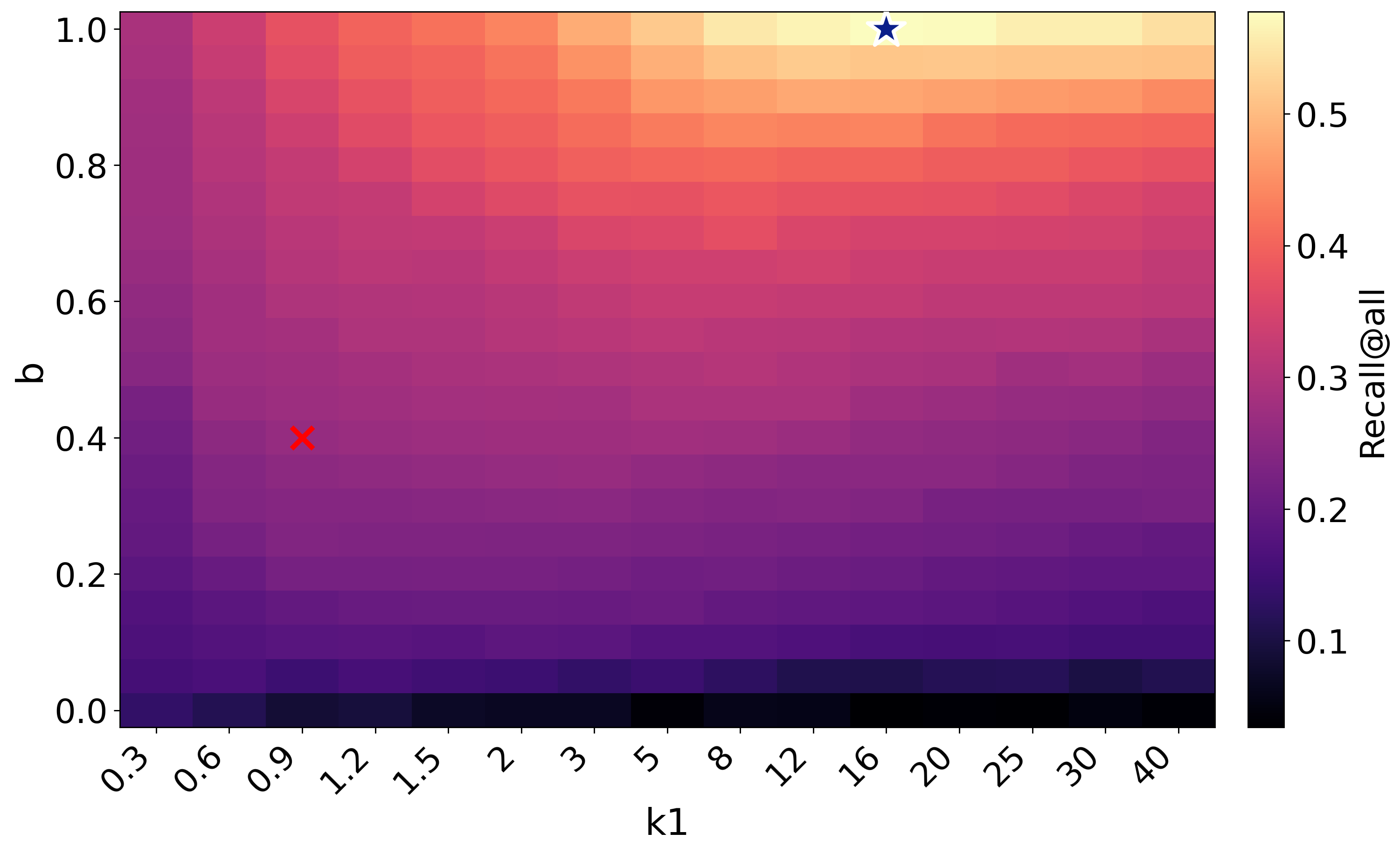}
\caption{Grid search results of BM25 tuning over different $k_1$ and $b$ combinations. \textcolor{blue}{$\bigstar$} indicates the best-performing configuration, while \textcolor{red}{$\times$} marks \textsc{Anserini}'s default setting: $(k_1=0.9, b=0.4)$.}
\label{fig:bm25-grid-search}
\end{figure}

We sample 100 queries from BrowseComp-Plus and perform a grid search over $k_1$ and $b$.
Figure~\ref{fig:bm25-grid-search} shows that \textsc{Anserini}'s default setting $(k_1=0.9, b=0.4)$ is placed in a low-performing region.
Stronger configurations use larger $k_1$ and higher $b$, with the best setting near $k_1=16$ and $b=1.0$.
This pattern suggests that the default BM25 setting is underfit for long-document evidence search.

The first and second columns in Table~\ref{tab:ablation-bm25-tuning} show the downstream effect on a 100-query BrowseComp-Plus subset.
The default setting uses $k_1=0.9$ and $b=0.4$, while the tuned setting uses $k_1=25$ and $b=1$, following Table~\ref{tab:main-results}.
We also tested the best grid-search setting, $k_1=16$ and $b=1$, and observed similar results.
Tuning raises accuracy from 64.0\% to 82.0\% (81.1\% with the best grid-search setting) and surfaced recall from 84.6\% to 95.7\% (94.0\% with the best grid-search setting).
Previewed recall and behavior recall also improve, showing that better first-stage ranking gives the agent more useful evidence to inspect.

This sensitivity is unsurprising given the length of BrowseComp-Plus documents: the median document has $\sim$2k tokens, and the 90th percentile has $\sim$14k tokens (cf. Figure~4 in~\citet{chen2025browsecompplusfairtransparentevaluation}).
At this length, relevant evidence is often buried in irrelevant text, making length normalization and term-frequency saturation matter more than in passage retrieval.
\textsc{Anserini}'s default BM25 setting is tuned for that shorter-document regime, whereas tuned parameters better match long-document evidence search.
This adaptation improves both evidence recall and final answer accuracy, consistent with the prior findings~\cite{meng2026revisiting}.

\paragraph{Cost-Efficiency}
\label{abl:cost-efficiency}
To better understand which designs in \textsc{Pi-Serini} improve cost efficiency, we focus on three factors: (1) BM25 tuning, (2) termination policy, and (3) prefix cache, since these factors are the major differences between \textsc{Pi-Serini} and baseline systems.

\begin{table}[b]
\centering
\small
\resizebox{\linewidth}{!}{%
\begin{tabular}{lrrrr}
\toprule
\textbf{Metric} & \textbf{Default BM25} & \textbf{Tuned BM25} & \textbf{Tuned BM25} & \textbf{Tuned BM25} \\
BM25 $(k_1, b)$ & $(0.9, 0.4)$ & \multicolumn{3}{c}{$(25, 1)$} \\
Term. Condition & Timeout 300s & Timeout 300s & MaxIter100 & Timeout 3600s \\
\midrule
Accuracy (\%) & 64.0 & 82.0 & 76.0 & 83.0 \\
Acc.$^{\dagger}$ (\%) & 66.7 & 83.7 & 92.7 & 83.0 \\
\midrule
Surfaced (\%) & 84.6 & 95.7 & 97.3 & 96.8 \\
Previewed (\%) & 50.7 & 70.4 & 73.5 & 79.0 \\
Behavior (\%) & 36.4 & 52.2 & 56.5 & 60.0 \\
\midrule
Avg. Tool Calls & 22.6 & 23.1 & 27.3 & 27.1 \\
Med. Total Tokens & 281,324.0 & 271,140.5 & 263,880.5 & 218,085.0 \\
Med. Cached Tokens & 228,736.0 & 227,968.0 & 215,552.0 & 186,112.0 \\
\midrule
Cost (\$) & 24.1 & 21.7 & 24.9 & 26.3 \\
\bottomrule
\end{tabular}%
}
\caption{Performance, cost, and tool-use comparison between default/tuned BM25 under different termination policies tested by \textsc{Pi-Serini} with \code{gpt-5.4}.}
\label{tab:ablation-bm25-tuning}
\end{table}

We first examine the effect of BM25 tuning. The first and second columns of Table~\ref{tab:ablation-bm25-tuning} show that, under the same termination policy, cost can be slightly reduced with a better retriever. 
This phenomenon is consistent with the baseline results in Table~\ref{tab:main-results}, from row~\tabrow{main:o3-bm25} to row~\tabrow{main:gpt5-qwen3}.

For the termination policy, we compare three settings: (1) a 300-second timeout, which is the standard setting in \textsc{Pi-Serini}; (2) a maximum of 100 iterations, which aligns with the baseline setting~\cite{chen2025browsecompplusfairtransparentevaluation,meng2026revisiting}; and (3) a 3600-second timeout, which is an extremely large timeout used to simulate the absence of a practical termination policy. According to the second through last columns of Table~\ref{tab:ablation-bm25-tuning}, replacing the 100-iteration cap with the 300-second timeout slightly reduces cost, from \$24.9 to \$21.7, while substantially increasing the timeout to 3600 seconds does not dramatically increase the cost.

Regarding prefix cache, \textsc{Pi-Serini} inherits \textsc{Pi}'s agentic loop, which structures the interaction history and enables prefix-cache-friendly configuration by default. Although Table~\ref{tab:ablation-bm25-tuning} does not directly isolate the cost reduction from prefix caching, all settings exhibit high cached-token ratios: around 82\% to 90\% of total tokens are served from cache. Given the lower price of cached tokens, this suggests that the prefix-cache-friendly loop is a major contributor to \textsc{Pi-Serini}'s overall cost efficiency.

\section{Discussion}
\paragraph{Failure Mode Analysis.}
\label{dis:search_behavior}
GPT-5.5 and Claude Opus 4.7 are frontier models with comparable inference costs in our setting, yet they exhibit a substantial performance gap on BrowseComp-Plus.
Our trajectory analysis shows that both agents sometimes use internal knowledge or weakly related entities from retrieved documents to expand queries. 
A key behavioral difference is how the agent responds when such a probe is weak or incorrect. 
In cases where \code{gpt-5.5} outperforms \code{claude-opus-4.7}, we observe that \code{gpt-5.5} tends to keep candidate-specific probes reversible, returning to the original clues when they fail. 
Claude Opus 4.7, in contrast, often searches vertically within the same weak hypothesis. 
We refer to this failure mode as \emph{premature branch commitment}. 
Intuitively, if a query asks for a cozy ramen shop near Tokyo station with a red lantern outside, both agents may probe a familiar but incorrect candidate such as \textit{Ichiran}. If search agents cannot verify that \textit{Ichiran} matches the original clues, \code{gpt-5.5} would discard the hypothesis and return to the original clues, whereas \code{claude-opus-4.7} would continue issuing \textit{Ichiran}-related queries.
The real case from BrowseComp-Plus is shown in Appendix~\ref{app:query-trajectory}.

\paragraph{To Retrieve or Not To Retrieve.}
\label{retrieve_or_not}
Recent work by~\citet{cao2026codingagentseffectivelongcontext} and~\citet{li2026beyondsemantic} shows that coding agents can achieve strong performance on BrowseComp-Plus by reformulating deep research as a file-system navigation problem. 
In their setup, a coding agent interacts with a locally materialized subset of documents through command-line tools such as \texttt{rg}, \texttt{sed}, and \texttt{nl}, effectively reformulating a retrieval problem as a navigation problem via a file-based interface (cf. Figure~\ref{fig:accuracy-cost-tradeoff-loggrid} \code{codex w/ cli}, \code{DCI-lite}, and \code{DCI-CC}; we refer readers to their papers for further details).
However, this formulation assumes that the relevant search space can be localized in advance, which is difficult to satisfy in real-world settings where corpora are large, dynamic, and cannot be exhaustively stored or exposed.
Notably, in the experimental setup of~\citet{cao2026codingagentseffectivelongcontext}, documents are randomly sampled to construct a roughly 100k-token working set in order to fit within the context window. 
Because of these differences in problem formulation, and because researchers sometimes run subsampled evaluations to reduce cost~\cite{cao2026codingagentseffectivelongcontext}, we do not include a comparison in Table~\ref{tab:main-results} and report their results only as reference points in Figure~\ref{fig:accuracy-cost-tradeoff-loggrid}.

Nevertheless, their systems can be viewed as testing a setting where relevant documents are already available locally and the agent only needs to identify the needed context through simple interaction mechanisms.
Our results show a complementary point that, with proper configuration and sufficient retrieval depth, BM25 can already provide a high-recall candidate set for the agent to inspect.

\section{Conclusion}
We asked: ``Does a lexical retriever suffice as LLMs become more capable in an agentic loop?''
Our results suggest that, on BrowseComp-Plus, a well-configured lexical retriever with sufficient depth can suffice for effective deep research when paired with more capable LLMs in an agentic loop.
In our experiments, a well-configured lexical retriever with sufficient retrieval depth can match or exceed released dense-retriever baselines, while the prefix-cache-friendly agent harness makes full-benchmark runs far cheaper.
These findings suggest that the low scores of previously released BM25 baselines mainly stem from ill-configured parameters and shallow retrieval depth, rather than from an inherent failure of lexical retrieval.
The lesson is not that BM25 is enough everywhere, but that under-configured baselines can hide weak agent--retriever interaction and miss the opportunities in designing more effective and cost-efficient search agents.

For us as a community interested in continuing to push the boundaries of deep research and information-seeking systems, the most useful signal is where \textsc{Pi-Serini} fails.
It often surfaces the right evidence documents, but the agent, with its pretrained policies and behaviors in this era, does not always browse, open, or use them effectively.
That failure mode shifts the question of system improvements from \emph{Can the retriever find the evidence?} to \emph{Can the agent recognize and spend context on the evidence it has already been given?}
Future progress in deep research may therefore come less from pulling the retriever lever harder, and more from designing tools that help agents navigate evidence documents with better judgments.

\section{Limitations}
\textsc{Pi-Serini} shows that a well-configured BM25 retriever, combined with a carefully designed tool interface, can surface high-recall evidence. 
However, our study has several limitations. 
First, \textsc{Pi-Serini} still relies on a relatively minimal interface for navigating the cached ranking, resulting in substantially lower previewed and behavior recall than surfaced recall. 
Second, our evaluation is limited to BrowseComp-Plus, leaving open the question of whether the current agent harness generalizes to other settings, including multilingual queries and domain-specific scenarios.
Finally, although using a timeout as the termination policy provides a simple way to control execution, queries with different levels of complexity may require different amounts of research time. 
A fixed time budget may therefore fail to balance effectiveness, latency, and monetary cost, leading either to insufficient exploration for difficult queries or unnecessary computation for simpler ones.

\bibliography{custom}

\clearpage
\onecolumn
\appendix

\section{System Setup}
\label{app:system-setup}

This appendix documents the exact \textsc{Pi-Serini} agent setup used in our experiments. We organize the setup by prompt, tool interface, backend configuration, and runtime policy. Placeholders in curly braces, such as \code{\{Question\}}, are replaced at runtime.\footnote{The setup documented in this appendix corresponds to \textsc{Pi-Serini}'s codebase at commit \texttt{68c5e0f}.}

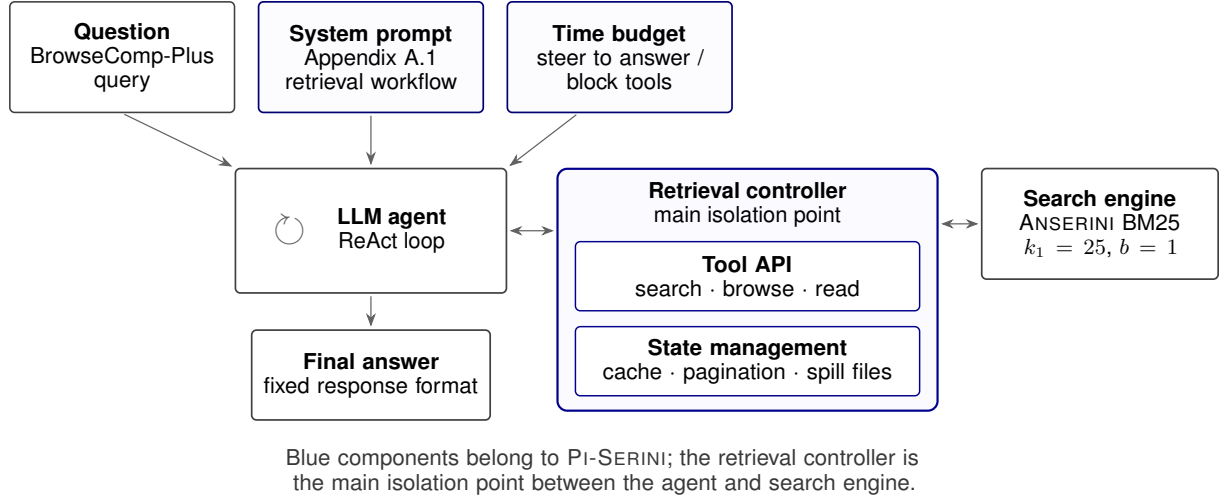
\begin{figure}[h]
\centering
\vspace{1mm}
\resizebox{\textwidth}{!}{%
\begin{tikzpicture}[
  font=\sffamily\small,
  >=Stealth,
  setup/.style={draw=black!75, fill=white, rounded corners=2pt, line width=0.65pt, align=center, text width=29mm, minimum height=16mm, inner sep=5pt},
  piconfig/.style={setup, draw=blue!45!black, fill=blue!1, line width=0.7pt},
  agent/.style={draw=black!75, fill=white, rounded corners=2pt, line width=0.7pt, align=center, text width=35mm, minimum height=18mm, inner sep=6pt},
  answer/.style={draw=black!75, fill=white, rounded corners=2pt, line width=0.7pt, align=center, text width=31mm, minimum height=13mm, inner sep=5pt},
  controller/.style={draw=blue!55!black, fill=blue!2, rounded corners=4pt, line width=0.9pt, align=center, minimum width=55mm, minimum height=35mm, inner sep=6pt},
  component/.style={draw=blue!55!black, fill=white, rounded corners=2pt, line width=0.6pt, align=center, text width=47mm, minimum width=48mm, minimum height=10mm, inner sep=4pt},
  engine/.style={draw=black!75, fill=white, rounded corners=2pt, line width=0.7pt, align=center, text width=30mm, minimum height=16mm, inner sep=6pt},
  arrow/.style={-{Stealth[length=2.0mm,width=1.45mm]}, line width=0.5pt, draw=black!62, shorten <=1pt, shorten >=1pt},
  bidir/.style={<->, line width=0.5pt, draw=black!62, shorten <=1pt, shorten >=1pt}
]

\node[setup] (question) at (-5.6,2.55) {\textbf{Question}\\BrowseComp-Plus\\query};
\node[piconfig] (prompt) at (-2.0,2.55) {\textbf{System prompt}\\Appendix A.1\\retrieval workflow};
\node[piconfig] (budget) at (1.6,2.55) {\textbf{Time budget}\\steer to answer /\\block tools};

\node[agent, anchor=north] (agent) at (-2.0,0.95) {};
\node[align=center, font=\sffamily\small] at ([xshift=3.0mm]agent.center) {\textbf{LLM agent}\\ReAct loop};
\node[font=\LARGE, text=black!62] at ([xshift=-11.8mm]agent.center) {$\circlearrowright$};
\node[answer] (answer) at (-2.0,-2.05) {\textbf{Final answer}\\fixed response format};

\node[controller, anchor=north] (controller) at (3.45,0.95) {};
\node[align=center, font=\sffamily\small] at (3.45,0.42) {\textbf{Retrieval controller}\\main isolation point};
\node[component] (toolapi) at (3.45,-0.62) {\textbf{Tool API}\\search $\cdot$ browse $\cdot$ read};
\node[component] (state) at (3.45,-1.85) {\textbf{State management}\\cache $\cdot$ pagination $\cdot$ spill files};
\node[engine, anchor=north] (engine) at (8.55,0.95) {\textbf{Search engine}\\\textsc{Anserini} BM25\\$k_1=25$, $b=1$};

\draw[arrow] (question.south) -- (agent.north west);
\draw[arrow] (prompt.south) -- (agent.north);
\draw[arrow] ([xshift=-6mm]budget.south) -- (agent.north east);
\draw[bidir] (agent.east) -- (controller.west |- agent.east);
\draw[bidir] (controller.east |- engine.west) -- (engine.west);
\draw[arrow] (agent.south) -- (answer.north);

\node[align=center, font=\sffamily\small, text=black!78, text width=130mm] at (1.35,-3.45)
{Blue components belong to \textsc{Pi-Serini}; the retrieval controller is the main isolation point between the agent and search engine.};

\end{tikzpicture}%
}
\vspace{1mm}
\par\noindent\makebox[\textwidth]{\rule{0.96\textwidth}{0.4pt}}
\vspace{-1mm}
\caption{System architecture of \textsc{Pi-Serini}. The system includes the retrieval workflow prompt, time-budget steering policy, and retrieval controller. The LLM agent interacts with \textsc{Anserini} only through the controller, which exposes a constrained tool API and maintains local retrieval state for controlled, paginated BM25 retrieval.}
\label{fig:appendix-piserini-system}
\end{figure}

\subsection{Agent Prompt}
\label{app:agent-prompt}

\begin{appendixsetupbox}{Search Agent Prompt}
\begin{Verbatim}[fontsize=\scriptsize,breaklines=true,breakanywhere=true,breaksymbolleft={},breaksymbolright={}]
You are a deep research agent answering a question using only the provided tools.

Workflow:
1. Use search with a concise raw query string based on the original question.
2. Prefer short lexical searches over long natural-language rewrites.
3. Browse the current ranking with read_search_results before repeatedly rewriting the query.
4. If a promising candidate document appears in the ranking, inspect it with read_document.
5. When reading a document, start with offset=1 and a moderate limit. If it is truncated and still relevant, continue reading the same document.
6. Use search refinements only when they add a genuinely new clue from what you already saw.
7. Every call to search, read_search_results, and read_document must include reason as the first argument. Keep it specific, under 100 words, and focused on the clue, gap, candidate, or ranking issue.
8. As soon as you have enough evidence, stop using tools and answer in plain assistant text.
9. Your final response must use exactly this format:
   Explanation: {your explanation for your final answer. Cite supporting docids inline in square brackets [] at the end of sentences when possible, for example [123].}
   Exact Answer: {your succinct, final answer}
   Confidence: {your confidence score between 0% and 100%}
10. If you later receive a user steer telling you to submit now, stop using tools immediately and answer right away with the exact final response format below. Do not do more research after that steer.
11. Keep Exact Answer concise and directly responsive to the question.

Question: {Question}
\end{Verbatim}
\end{appendixsetupbox}

\subsection{Retrieval Tool Interface}
\label{app:retrieval-tools}

\begin{appendixsetupbox}{Tool: search}
\begin{Verbatim}[fontsize=\scriptsize,breaklines=true,breakanywhere=true,breaksymbolleft={},breaksymbolright={}]
Purpose: Send a raw lexical query to the configured retrieval backend.
Arguments:
  reason: brief rationale, supplied first, at most 100 words
  query: raw query string; not a structured object and not raw Lucene syntax
Runtime behavior in our BM25 runs:
  query_mode = plain
  backend request limit = 1000 hits
  initial displayed page = ranks 1-5
Output:
  search_id, formatted first page, cached docids, displayed docids, timing/truncation metadata
\end{Verbatim}
\end{appendixsetupbox}

\begin{appendixsetupbox}{Tool: read\_search\_results}
\begin{Verbatim}[fontsize=\scriptsize,breaklines=true,breakanywhere=true,breaksymbolleft={},breaksymbolright={}]
Purpose: Browse an existing cached ranking by search_id without issuing a new backend query.
Arguments:
  reason: brief rationale, supplied first, at most 100 words
  search_id: identifier returned by search
  offset: optional 1-indexed rank offset; default is 6
  limit: optional number of ranked hits to show; default is 10
Output:
  formatted page from the cached ranking, displayed docids, next_offset when more hits remain,
  and truncation metadata if the formatted page is too large
\end{Verbatim}
\end{appendixsetupbox}

\begin{appendixsetupbox}{Tool: read\_document}
\begin{Verbatim}[fontsize=\scriptsize,breaklines=true,breakanywhere=true,breaksymbolleft={},breaksymbolright={}]
Purpose: Read one backend document by docid in line-based chunks.
Arguments:
  reason: brief rationale, supplied first, at most 100 words
  docid: document identifier from search or read_search_results
  offset: optional 1-indexed line offset; default is 1
  limit: optional maximum number of lines; default is 200
Output for a found document:
  formatted document lines, returned line range, total line count, backend truncation flag,
  next_offset when available, timing metadata, and output-truncation metadata when applicable
\end{Verbatim}
\end{appendixsetupbox}

\subsection{Backend and Cache Configuration}
\label{app:backend-cache-config}

\begin{appendixsetupbox}{Retrieval Backend Configuration}
\begin{Verbatim}[fontsize=\scriptsize,breaklines=true,breakanywhere=true,breaksymbolleft={},breaksymbolright={}]
Experimental backend used in this paper:
  retriever = \textsc{Anserini} BM25
  corpus = BrowseComp-Plus
  BM25 parameters = k1 25, b 1

pi-search configuration contract:
  PI_SEARCH_EXTENSION_CONFIG is required and must be supplied by the caller
  our experiments use \textsc{Anserini}'s BM25 backend kind

Backend kinds supported by the pi-search config schema:
  - \textsc{Anserini}-bm25 over TCP: {backend: {kind: "\textsc{Anserini}-bm25", transport: {kind: "tcp", host, port}}}
  - \textsc{Anserini}-bm25 over stdio: {backend: {kind: "\textsc{Anserini}-bm25", transport: {kind: "stdio", indexPath}}}
  - http-json: {backend: {kind: "http-json", capabilities, endpoints}}
  - mock: {backend: {kind: "mock", documents}}
\end{Verbatim}
\end{appendixsetupbox}

\begin{appendixsetupbox}{Search Cache, Pagination, and Spill Files}
\begin{Verbatim}[fontsize=\scriptsize,breaklines=true,breakanywhere=true,breaksymbolleft={},breaksymbolright={}]
Search cache:
  each search call receives a session-local search_id such as s1, s2, ...
  each search_id stores the raw query, query mode, and retrieved ranking
  at most 32 search_id entries are retained; older entries are evicted first

Ranking depth and pagination:
  search requests up to 1000 backend hits and caches the returned ranking
  search displays only ranks 1-5 immediately
  read_search_results defaults to offset=6 and limit=10 when omitted
  read_document defaults to offset=1 and limit=200 lines when omitted

Spill-file behavior:
  tool output is first formatted for the agent context
  if formatted output exceeds Pi's line or byte truncation limits, the visible
  output is truncated and the complete search page or document chunk is saved
  under a temporary pi-search spill directory
  the temporary spill directory is cleaned up on session shutdown or process exit
\end{Verbatim}
\end{appendixsetupbox}

\subsection{Time-Budget Policy}
\label{app:time-budget-policy}

\begin{appendixsetupbox}{Submit-Now Steering Policy}
\begin{Verbatim}[fontsize=\scriptsize,breaklines=true,breakanywhere=true,breaksymbolleft={},breaksymbolright={}]
Per-query timeout in the main experiments: T = 300 seconds
Activation condition:
  the timer is enabled only when TIMEOUT_SECONDS is set to a positive finite value
Submit-now trigger:
  delay = floor(0.7 * TIMEOUT_SECONDS * 1000) milliseconds
Submit-now behavior when the timer fires while the agent is active:
  1. Send a user steer saying the time budget is nearly exhausted.
  2. Instruct the agent to stop using tools and submit its best answer immediately.
  3. Block later calls to search, read_search_results, and read_document.
Final response format after the steer remains:
  Explanation: ...
  Exact Answer: ...
  Confidence: ...
\end{Verbatim}
\end{appendixsetupbox}

\subsection{Gold-Answer LLM Judge}
\label{app:gold-answer-judge}

We evaluate final-answer correctness with a gold-answer LLM judge.
The judge receives the question, the agent's final response, and the benchmark-provided correct answer.
The default judge model is \code{openai-codex/gpt-5.3-codex}, run through \textsc{Pi} in JSON mode.
Each judge call uses an isolated \code{PI\_CODING\_AGENT\_DIR}, writes raw judge events and stderr logs, and times out after 180 seconds.

\begin{appendixsetupbox}{Gold-Answer Judge Configuration}
\begin{Verbatim}[fontsize=\scriptsize,breaklines=true,breakanywhere=true,breaksymbolleft={},breaksymbolright={}]
Pi invocation:
  pi --no-tools --no-session --no-skills --mode json \
     --model openai-codex/gpt-5.3-codex --thinking low {judge_prompt}
Judge timeout: 180 seconds per query
Evaluation output root: evals/pi_judge
Ground-truth file:
  data/browsecomp-plus/ground-truth/browsecomp_plus_decrypted.jsonl
Per-query outputs:
  per-query/{query_id}_eval.json
  raw-events/{query_id}.jsonl
  stderr/{query_id}.log
\end{Verbatim}
\end{appendixsetupbox}

\begin{appendixsetupbox}{Gold-Answer Judge Prompt}
\begin{Verbatim}[fontsize=\scriptsize,breaklines=true,breakanywhere=true,breaksymbolleft={},breaksymbolright={}]
You are an evaluation judge.

Your job is to determine whether the response's final answer is semantically equivalent to the known correct answer.
Do not solve the question yourself.
Do not use outside knowledge.
Focus only on whether the response's final answer matches the correct answer.
Allow harmless wording differences, equivalent formatting, and added correct detail.
For numerical answers, allow small formatting differences and obvious equivalent forms.
If the response does not contain a final answer you can extract, set extracted_final_answer to null and correct to false.

Return exactly one JSON object and nothing else.
Do not wrap the JSON in markdown or code fences.
Use this exact schema:
{
  "extracted_final_answer": string | null,
  "correct_answer": string,
  "reasoning": string,
  "correct": boolean,
  "confidence": number
}

Requirements:
- confidence must be a number between 0 and 100
- correct must be true or false
- repeat the provided correct answer exactly in correct_answer
- reasoning must explain only whether the extracted final answer matches the correct answer

Question: {question}

Response:
{response}

Correct answer: {correct_answer}
\end{Verbatim}
\end{appendixsetupbox}

\section{Model Token Pricing Used in Benchmark Runs}
\label{app:token-pricing}

Table~\ref{tab:token-prices} reports the token pricing used by our benchmark runtime.
All prices are in \textbf{USD per 1M tokens}.

\begin{table}[h]
\centering
\caption{Token pricing used during benchmark execution. All prices are in USD per 1M tokens.}
\label{tab:token-prices}
\small
\setlength{\tabcolsep}{7pt}
\renewcommand{\arraystretch}{1.08}
\begin{tabular}{lrrr}
\toprule
\textbf{Model} & \textbf{Input} & \textbf{Output} & \textbf{Cache Read} \\
\midrule
\code{claude-haiku-4.5}  & 1.00 & 5.00  & 0.10 \\
\code{claude-opus-4.7}   & 5.00 & 25.00 & 0.50 \\
\code{gpt-5}             & 1.25 & 10.00 & 0.125 \\
\code{gpt-5.2}           & 1.75 & 14.00 & 0.175 \\
\code{gpt-5.3-codex}     & 1.75 & 14.00 & 0.175 \\
\code{gpt-5.4-mini}      & 0.75 & 4.50  & 0.075 \\
\code{gpt-5.4}           & 2.50 & 15.00 & 0.25 \\
\code{gpt-5.5}           & 5.00 & 30.00 & 0.50 \\
\code{deepseek-v4-flash} & 0.14 & 0.28  & 0.028 \\
\code{deepseek-v4-pro}   & 1.74 & 3.48  & 0.145 \\
\bottomrule
\end{tabular}
\end{table}

\section{Premature Branch Commitment}
\label{app:query-trajectory}
This example shows the query trajectories for Query 678 and demonstrates that incorrect query expansion is not inherently fatal. GPT-5.5 also introduces unverified candidate terms, such as \textcolor{orange}{Warrington}, \textcolor{orange}{Vinegar Strokes}, and \textcolor{orange}{Arts Educational}, which lead to incorrect search directions. However, these probes remain reversible: after unproductive searches, GPT-5.5 returns to the root constraints, such as the town population and spelling-history clue, and then follows the evidence-supported ``Crawley'' path. This eventually leads to the correct candidate, ``Dani Sylvia,'' and the query is refined using the original interview constraints.

Claude Opus 4.7 follows a different pattern. After the first search retrieves a title mentioning \textcolor{orange}{Renato Paris} and \textcolor{orange}{CICAS}, the agent treats this superficial match as a candidate anchor. It then searches vertically within the Renato Paris branch, adding more unverified terms such as \textcolor{orange}{Tom Misch}, \textcolor{orange}{Italy}, \textcolor{orange}{Forl\`i}, \textcolor{orange}{ICMP}, \textcolor{orange}{BIMM}, and \textcolor{orange}{Berklee}.
These queries deepen the same hypothesis rather than testing alternative interpretations of the original clues. The issue is therefore not that Claude uses domain knowledge, but that it keeps elaborating a weak branch after the supporting evidence is insufficient.

\begin{appendixsetupbox}{Original Query}
\begin{Verbatim}[
  fontsize=\scriptsize,
  breaklines=true,
  breaksymbolleft={},
  breaksymbolright={}
]
Find the school mentioned in a November 2019 interview by an artist who grew up in a town with 2019 population around 114,000 and current spelling appearing around 1316; performed two roles in the same industry; won a late-2010s award; released a creation in 2019; described their style as "eclectic"; described social media as "dangerous"; and said Instagram or Twitter were the best way to connect.
\end{Verbatim}
\end{appendixsetupbox}

\begin{appendixsetupbox}{GPT-5.5 Query Trajectory}
\begin{Verbatim}[fontsize=\tiny, breaklines=false, commandchars=\\\{\}]
Root: Original query

|-- Step 1
|   Reason: Locate the November 2019 interview containing the exact quoted style/social media phrases.
|   Query:  "eclectic" "social media is dangerous" "best way to connect" interview November 2019 artist
|
|-- Step 2
|   Reason: Search exact dangerous quote with November interview lexical clues, dropping broader social phrase.
|   Query:  "social media is dangerous" "Instagram" "Twitter" "interview" "2019" artist
|
|-- Step 3
|   Reason: Find pages with both eclectic self-description and contact via social platforms from music interviews.
|   Query:  "How would you describe your style" "eclectic" "social media" "dangerous"
|
|-- Step 4
|   Reason: Search exact interview question about style with release-year clue likely in independent artist Q&A.
|   Query:  "describe your style as eclectic" "interview" "2019" musician
|
|-- Step 5
|   Reason: Target the requested answer phrase in the interview: artist says social media dangerous and mentions training school.
|   Query:  "social media" "dangerous" "trained at" "eclectic" "interview"
|
|-- Step 6
|   Reason: Exact dangerous phrase alone may reveal the interview despite noisy broad results.
|   Query:  "social media is dangerous"
|
|-- Step 7
|   Reason: Look for variant wording around social media danger plus eclectic in artist interviews.
|   Query:  "social media" "dangerous" "eclectic" "Instagram" "interview" "November"
|
|-- Step 8
|   Reason: Use the town clue Warrington plus exact social-media/interview terms to identify the artist.
|   Query:  "grew up in \textcolor{orange}{Warrington}" artist "eclectic" "social media" "dangerous"
|   |
|   `-- Step 9
|       Reason: Search Warrington artist and 2019 interview clue without exact dangerous phrase.
|       Query:  "grew up in \textcolor{orange}{Warrington}" "interview" "eclectic" "2019"
|
`-- Step 10
    Reason: Identify town from population 114,000 and name-spelling clue.
    Query:  "current spelling" "first appeared" "1316" town population 114000
    |
    |-- Step 11
    |   Reason: Use identified town Crawley with the interview quote to find the artist.
    |   Query:  "grew up in Crawley" "social media" "dangerous" "eclectic"
    |
    |-- Step 12
    |   Reason: Search Crawley plus November 2019 interview and eclectic without exact grew-up phrase.
    |   Query:  "Crawley" "eclectic" "social media" "interview" "November" "2019" artist
    |
    |-- Step 13
    |   Reason: Candidate from Crawley clue: Vinegar Strokes; verify interview wording and training school.
    |   Query:  "\textcolor{orange}{Vinegar Strokes}" "eclectic" "social media" "dangerous" "trained"
    |   |
    |   `-- Step 14
    |       Reason: Find candidate biography documents for Vinegar Strokes and relevant facts.
    |       Query:  "\textcolor{orange}{Vinegar Strokes}" "grew up in Crawley" "\textcolor{orange}{Arts Educational}"
    |
    `-- Step 15
        Reason: Candidate Dani Sylvia matches Crawley, 2019 release, songwriter award; locate November 2019 interview with school answer.
        Query:  "Dani Sylvia" "eclectic" "social media" "dangerous" "school"
\end{Verbatim}
\end{appendixsetupbox}

\begin{appendixsetupbox}{Claude Opus 4.7 Query Trajectory}
\begin{Verbatim}[fontsize=\tiny, breaklines=false, commandchars=\\\{\}]
Root: Original query

|-- Step 1
|   Reason: Find the November 2019 interview where artist calls style "eclectic" and social media "dangerous".
|   Query:  interview November 2019 style "eclectic" social media "dangerous"
|   |
|   `-- Step 2
|       Reason: Find the actual Renato Paris CICAS interview content.
|       Query:  \textcolor{orange}{Renato Paris} \textcolor{orange}{CICAS} interview eclectic dangerous
|       |
|       |-- Step 4
|       |   Reason: Find info about Renato Paris artist hometown school training.
|       |   Query:  \textcolor{orange}{Renato Paris} musician singer producer hometown
|       |
|       |-- Step 5
|       |   Reason: CICAS podcast about Renato Paris.
|       |   Query:  \textcolor{orange}{CICAS} podcast \textcolor{orange}{Renato Paris} keyboardist
|       |
|       |-- Step 6
|       |   Reason: Find Renato Paris biography or interview content elsewhere in corpus.
|       |   Query:  \textcolor{orange}{Renato Paris} interview eclectic style
|       |
|       |-- Step 7
|       |   Reason: Search for Renato Paris bio---he might be a different artist.
|       |   Query:  "\textcolor{orange}{Renato Paris}" biography born
|       |
|       |-- Step 10
|       |   Reason: Renato Paris Tom Misch keyboardist Forli.
|       |   Query:  \textcolor{orange}{Renato Paris} \textcolor{orange}{Tom Misch} keyboards \textcolor{orange}{Italy}
|       |   |
|       |   `-- Step 11
|       |       Reason: Forlì Italy population history spelling.
|       |       Query:  \textcolor{orange}{Forlì} Wikipedia population history name origin
|       |       |
|       |       |-- Step 12
|       |       |   Reason: Look for Forlì Wikipedia article in the corpus.
|       |       |   Query:  \textcolor{orange}{Forlì} city Italy Emilia-Romagna
|       |       |
|       |       |-- Step 16
|       |       |   Reason: Try different angle---artist from Forlì grew up there.
|       |       |   Query:  "grew up in \textcolor{orange}{Forlì}" musician artist
|       |       |
|       |       `-- Step 17
|       |           Reason: Try searching for transcription of the interview content with Italian connection.
|       |           Query:  \textcolor{orange}{Renato Paris} singer Italian \textcolor{orange}{Forlì} conservatory
|       |
|       |-- Step 13
|       |   Reason: Search general corpus for Renato Paris music school training.
|       |   Query:  \textcolor{orange}{Renato Paris} trained school music London
|       |
|       |-- Step 15
|       |   Reason: CICAS abbreviation context---what is it.
|       |   Query:  \textcolor{orange}{CICAS} podcast meaning organization music creative
|       |
|       `-- Step 18
|           Reason: Look for Renato Paris ICMP or BIMM or Tech Music School.
|           Query:  "\textcolor{orange}{Renato Paris}" \textcolor{orange}{ICMP} \textcolor{orange}{BIMM} \textcolor{orange}{Tech Music School} \textcolor{orange}{Berklee} \textcolor{orange}{Trinity}
|
|-- Step 8
|   Reason: Try finding the exact interview piece via different phrasing.
|   Query:  "my style is eclectic" interview "dangerous" Instagram Facebook
|
|-- Step 9
|   Reason: Search for specific quoted phrases from a November 2019 artist interview.
|   Query:  "social media is dangerous" interview 2019 musician eclectic
|
`-- Step 14
    Reason: Try finding interview with specific phrases about connecting via social media.
    Query:  "best way to connect" Instagram Twitter interview eclectic 2019
\end{Verbatim}
\end{appendixsetupbox}

\end{document}